\documentclass[aps,pra,showpacs,twocolumn,floatfix]{revtex4}
\usepackage{graphicx} 
\usepackage{amsmath}
\usepackage{bm}
\usepackage{dcolumn}

\begin{document}

\title{Relativistic many-body calculations of the Stark-induced
amplitude of the $6P_{1/2}\ -7P_{1/2}$ transition   in thallium}

\author{M. S. Safronova}
\email{msafrono@udel.edu} \affiliation {Department of Physics and
Astronomy, 223 Sharp Lab,
 University of Delaware, Newark, Delaware 19716}

\author{W. R. Johnson}
\email{johnson@nd.edu} \homepage{www.nd.edu/~johnson}
\affiliation{Department of Physics,  University of Notre Dame,
Notre Dame, Indiana 46556}

\author{ U. I. Safronova }
\email{usafrono@nd.edu}\altaffiliation{ On  leave  from ISAN,
Troitsk, Russia} \affiliation{Physics Department, University of
Nevada, Reno, Nevada
 89557}

\author{ T. E. Cowan }
\affiliation{Physics Department, University of Nevada, Reno, Nevada
 89557}

\date{\today}
\begin{abstract}
Stark-induced amplitudes for the $6P_{1/2} - 7P_{1/2}$ transition
in Tl~I are calculated using the relativistic SD approximation in
which single and double excitations of Dirac-Hartree-Fock
levels are summed to all orders in perturbation theory.  Our SD
values $\alpha_S$ = 368~$a_0^3$ and $\left|\beta_S\right|$= 298~$a_0^3$ are in good
agreement with the measurements $\alpha_S$=377(8)~$a_0^3$ and $\beta_S$ =
313(8)~$a_0^3$  by D. DeMille, D. Budker, and E. D. Commins [Phys.\
Rev.\ A{\bf 50}, 4657 (1994)]. Calculations of the
Stark shifts in the  $6P_{1/2} - 7P_{1/2}$ and  $6P_{1/2} - 7S_{1/2}$
transitions are also carried out.  The Stark shifts
predicted by our calculations agree with the most accurate measured values
within the experimental uncertainties for both transitions.

\end{abstract}
\maketitle

\section{Introduction}
The 293 nm $6P_{1/2}-7P_{1/2}$ transition in thallium has been
studied extensively both experimentally and theoretically
because of its connection to atomic parity nonconservation (PNC).
Although $6P_{1/2}-7P_{1/2}$ is nominally a magnetic dipole (M1)
transition, there is also an electric dipole (E1) component
arising from the weak interaction  mediated by $Z_0$ exchange
between the nucleus and bound electrons. The weak E1 component of
the $6P_{1/2}-7P_{1/2}$ transition is calibrated using the
Stark-induced amplitude $\beta_S$. Measured values of both
$\alpha_S$~=~2.01(4)$\times$10$^{-5}\mu_B$~cm/V and
$\beta_S$~=~1.67(4)$\times$10$^{-5}\mu_B$~cm/V for the
$6P_{1/2}-7P_{1/2}$ transition in thallium were reported by
\citet{st-94} and found to be in substantial disagreement with
earlier measurements $\alpha_S$~=~1.31(6)$\times$10$^{-5}\mu_B$~cm/V
and $\beta_S$~=~1.09(5)$\times$10$^{-5}\mu_B$~cm/V by \citet{st-86}.
It was pointed out in Ref.~\cite{st-02} that  both the Stark shift
and the Stark-induced amplitude could be measured in the same
transition in which PNC is measured \cite{pnc-95}.

High-precision measurements of the Stark shift within the 378~nm
$6P_{1/2}-7S_{1/2}$  E1 transition in atomic thallium were
recently reported by \citet{st-02}.  The result $\Delta \nu_S$ =
-103.23(39)~kHz/(kV/cm)$^2$ had higher accuracy by
factor of 15 than earlier measurements \cite{st-94,st-70}.
The earlier value for the
$6P_{1/2}-7S_{1/2}$ transition from Ref.~\cite{st-94} was
$\Delta \nu_S$~=~-112(6)~kHz/(kV/cm)$^2$.

In the present paper, we carry out relativistic all-order calculations of
Stark-induced amplitudes
$\alpha_S$ and $\beta_S$ for the $6P_{1/2}-7P_{1/2}$ transition
as well as  Stark shifts within both $6P_{1/2}-7S_{1/2}$  and
$6P_{1/2}-7P_{1/2}$ transitions  in
atomic thallium. The  calculations are carried out using
the relativistic SD all-order method in which single and double excitations
of Dirac-Hartree-Fock (DHF) wave functions are summed to all orders in perturbation
theory. Recently, lifetimes, E1, E2, and M1 transition rates,
hyperfine constants, and excitation energies of the $nP_{J}$,
$nS_{1/2}$, and $nD_{J}$ states in neutral thallium were evaluated
by \citet{tl-05}, using both relativistic many-body perturbation
theory (MBPT) and the relativistic SD method. The SD calculations were
found to be in excellent agreement with the best available
experimental data.

\section{Stark-induced amplitude of the $6P_{1/2}\ -7P_{1/2}$ transition
  in thallium}

Stark-induced scalar and vector polarizabilities $\alpha_S$ and
$\beta_S$ for transitions in Na, K, Rb, Cs, and Fr were calculated
in the SD approximation by \citet{safr-alk}.  Following the
procedure used in \cite{tl-05}, we treat Tl~I as a
one-valence-electron atom with a Hg-like core and evaluate the
Stark-induced scalar and vector polarizabilities $\alpha_S$ and
$\beta_S$ following the procedures used in \cite{safr-alk} for
alkali-metal atoms.

\begin{table*}
\caption{\label{tab-comp1}  The contributions to Stark-induced
scalar polarizability $\alpha_S$ for the $6p_{1/2}-7p_{1/2}$ transition in Tl~I. The
corresponding energy differences
and electric-dipole reduced matrix elements are also listed. All values are given in a.u.}
\begin{ruledtabular}
\begin{tabular}{lrrrrrr}
\multicolumn{1}{c}{Contribution}&
\multicolumn{1}{c}{$nlj$}& \multicolumn{1}{c}{$E_{6P_{1/2}}-E_{nlj}$ }&
\multicolumn{1}{c}{$E_{7P_{1/2}}-E_{nlj}$ }& \multicolumn{1}{c}{$Z_{7P_{1/2},nlj}$}&
\multicolumn{1}{c}{$Z_{nlj,6P_{1/2}}$ }&
\multicolumn{1}{c}{$\alpha_S$ \vspace{0.1cm}}\\
\hline
 $\alpha^{\text{main}}_S(nS)$&  $7S_{1/2}$&  -0.120640&    0.035004&       6.016&  -1.827&  -37.2    \\
                      & $8S_{1/2}$&  -0.176539&   -0.020895&      -6.215&  -0.535&  -29.7      \\
                      & $9S_{1/2}$&  -0.196680&   -0.041036&      -1.274&  -0.298&   -1.9      \\
                      & $10S_{1/2}$& -0.206387&   -0.050743&      -0.656&  -0.200&   -0.5      \\
 $\alpha^{\text{tail}}_S(nS)$&           &           &            &     &    &   -2.0  \\[0.5pc]
 $\alpha^{\text{main}}_S(nD_{3/2})$& $6D_{3/2}$&  -0.164565&   -0.008921&     -10.703&  -2.334&   492.0   \\
                      & $7D_{3/2}$&  -0.191418&   -0.035774&       4.821&  -1.101&   -29.4   \\
                      & $8D_{3/2}$&  -0.203543&   -0.047899&       2.377&  -0.672&    -6.9  \\
                      & $9D_{3/2}$&  -0.210040&   -0.054396&       1.535&  -0.476&    -2.8   \\
  $\alpha^{\text{tail}}_S(nD_{3/2})$&&     &            &            &        &   -14.1 \\[0.5pc]
  $\delta\alpha^{\text{core}}_S$             &     &         &   &            &        &     0.4 \\[0.5pc]
  Total                & &            &            &        &   & 368  \\
\end{tabular}
\end{ruledtabular}
\end{table*}

The scalar and vector polarizabilities $\alpha_S$ and $\beta_S$
for the transitions between the ground state $6P_{1/2}$ and the
excited state $7P_{1/2}$ in Tl~I are calculated as
\begin{eqnarray}\label{eq1}
\alpha_{{\rm S}} &=&
\sum_{n}\left[I_{\alpha}(nS_{1/2})-I_{\alpha }(nD_{3/2})\right]\, ,  \\ \label{eq12}
\beta _{{\rm S}}&=&
\sum_{n}\left[I_{\beta}(nS_{1/2})+\frac{1}{2}I_{\beta }(nD_{3/2})\right]\,
,
\end{eqnarray}
where
\begin{eqnarray*}
\lefteqn{I_{\alpha}(nlj)}\hspace{0em} \nonumber  \\
&&=\frac{1}{6}
\left[ \frac{Z_{7P_{1/2},nlj} Z_{nlj,6P_{1/2}}}
{E_{6P_{1/2}}-E_{nlj}}+
\frac{Z_{7P_{1/2},nlj} Z_{nlj,6P_{1/2}}}
{E_{7P_{1/2}}-E_{nlj}}\right]
\nonumber\,  \label{eq2}
\end{eqnarray*}
and
\begin{eqnarray*}
\lefteqn{I_{\beta }(nlj)}\hspace{0em} \nonumber  \\
&&=\frac{1}{6} \left[
\frac{Z_{7P_{1/2},nlj} Z_{nlj,6P_{1/2}}}
{E_{7P_{1/2}}-E_{nlj}}
-\frac{Z_{7P_{1/2},nlj} Z_{nlj,6P_{1/2}}}
{E_{6P_{1/2}}-E_{nlj}}\right] \nonumber\, .
\end{eqnarray*}
The quantities  $Z_{wv}$ in
the above equations are electric-dipole reduced matrix elements.
 The all-order SD matrix elements are
calculated as~\cite{tl-05}
\begin{equation}
Z_{wv}=\frac{z_{wv}+Z^{(a)}+ \cdots +Z^{(t)}}{\sqrt{(1+N_{w})(1+N_{v})}%
}\,,  \label{eq18}
\end{equation}
where $z_{wv}$ is the lowest-order (DHF) matrix element
and the terms $Z^{(k)}$, $k=a \cdots t$
are linear or quadratic function of the SD
excitation coefficients. The normalization terms $N_{w}$ are
quadratic functions of the excitation coefficients.
The SD matrix elements include all MBPT corrections through third order
together with important classes of forth- and higher-order corrections.

The calculation of the $\alpha _{\text{S}}$
is divided into three parts:
\begin{equation}
\alpha _{\text{S}}=
\alpha^{\text{main}}_S+\alpha^{\text{tail}}_S+\delta \alpha^{\text{core}}_S\,,  \label{main1}
\end{equation}
where $\alpha^{\text{main}}_S$ is the dominant contribution from
states near the valence state, $\delta\alpha^\text{core}_S$ is the
contribution from core-excited autoionizing states, and $\alpha^{\text{tail}}_S$ is the
remainder from highly excited one-electron states. Thus, we write,
\begin{eqnarray} \label{main2}
\alpha^{\text{main}}_S
&=&\sum_{n=7}^{10}I_{\alpha}(nS_{1/2})-%
\sum_{n=6}^{9}I_{\alpha}(nD_{3/2}),\     \\
\delta \alpha^{\text{core}}_S
&=&\sum_{n=1}^{6}I_{\alpha}(nS_{1/2})-\sum_{n=3}^{5} I_{\alpha}(nD_{3/2}),
\nonumber
\\
\alpha _{\text{tail}}
&=&\sum_{n=11}^{N}I_{\alpha}(nS_{1/2})-\sum_{n=10}^{N}I_{\alpha}(nD_{3/2})\, ,\nonumber
\end{eqnarray}
where $N$ is the number of the finite basis set states.
The calculation of $\beta_S$ is conducted in the same way.

We use B-splines \cite{Bspline} to generate a complete set of
DHF basis orbitals for use in the evaluation of all
electric-dipole matrix elements.  Here, we use here $N=50$ splines
(compared with 40 used in ~\cite{tl-05}) for each angular momentum.
The basis orbitals are constrained to a cavity of radius $R=90$
(a.u.).  The size of the cavity is taken to be large
enough to fit all of the states
needed for the calculation of the main terms for all of the
polarizabilities calculated in this work.
Furthermore, we use the Breit-Dirac-Hartree-Fock (BDHF) approximation here instead of the
Dirac-Hartree-Fock (DHF) approximation used in Ref.~\cite{tl-05} because we found that the
Breit contributes about 1\% to the values of $\alpha_S$ and $\beta_S$ at the DHF level. The
BDHF approximation includes the the one-body
part of the Breit interaction  in the DHF
equation (for more detail see Refs.~\cite{der-00,der-02}). The
one-body part of the Breit interaction is also included
in the equations that generate  DHF basis orbitals.
In fact, all of the calculations in this work, including the
calculations of the tail contributions,  are done with same
basis set.

The sums over $n$ in Eqs.~(\ref{eq1}, \ref{eq12}) converge rapidly; therefore,
only main term contributions have to be calculated accurately. We calculate main terms
for $\alpha_S$ and $\beta_S$  using SD matrix elements and experimental energies \cite{nist}.
The contribution of the remainder, while small, is not negligible
for $n<20$ and is calculated in the random-phase approximation (RPA) for
 these states. It is essentially zero for larger $n$ and is evaluated
 in the DHF approximation for $n \geq 20$ without loss of accuracy.
 We find that it is essential to use RPA approximation to
 evaluate the tail contribution for $\alpha_S$
  as DHF calculation significantly overestimates the tail yielding
  $-25~a^3_0$ while RPA calculation gives $-16~a^3_0$.
  DHF approximation is expected to significantly overestimate
  the value of the tail contribution as it significantly overestimates the value of the main
  term. The autoionizing contribution is very small, 0.1\%, for $\alpha_S$
and is negligible at the present level of accuracy for $\beta_S$.
It is evaluated in the DHF approximation.

\begin{table}
\caption{\label{tab-comp2}The contributions to Stark-induced
vector polarizability $\beta_S$(a.u.) for the $6P_{1/2}-7P_{1/2}$ transition
in Tl~I.}
\begin{ruledtabular}
\begin{tabular}{lrr}
\multicolumn{1}{c}{Contribution}&\multicolumn{1}{c}{$nlj$}&
\multicolumn{1}{c}{$\beta_S$ \vspace{0.1cm}}\\
\hline
 $\beta^{\text{main}}_S(nS)$&  $7S_{1/2}$& -67.5   \\
                      & $8S_{1/2}$&        -23.4   \\
                      & $9S_{1/2}$&        -1.2   \\
                      & $10S_{1/2}$&       -0.3   \\
 $\beta^{\text{tail}}_S(nS)$&&              -0.9   \\[0.5pc]
 $\beta^{\text{main}}_S(nD_{3/2})$&$6D_{3/2}$&  -220.7  \\
                      & $7D_{3/2}$&    10.1  \\
                      & $8D_{3/2}$&     2.1  \\
                      & $9D_{3/2}$&     0.8  \\
  $\beta^{\text{tail}}_S(nD_{3/2})$&&    3.3  \\[0.2pc]
  $\delta\beta^{\text{core}}_S$&&     0.0 \\[0.2pc]
  Total             &          &  -298 \\
\end{tabular}
\end{ruledtabular}
\end{table}

In Tables~\ref{tab-comp1} and \ref{tab-comp2}, we present the details
of our $\alpha_S$ and $\beta_S$  calculations for the transition
 between the ground state $6P_{1/2}$ and the excited  $7P_{1/2}$ state in Tl~I.
 We separate the contributions from the $nS$ and $nD_{3/2}$
 terms given by the Eqs.~(\ref{eq1}, \ref{eq12}).
 Furthermore, the contribution from each $n$ in the Eq.~(\ref{main2})
 for the main term
 is given separately in the corresponding $(nlj)$ row to demonstrate rapid convergence
 of the sums in Eqs.~(\ref{eq1}, \ref{eq12}). The tail
 contributions for $nS$ and $nD_{3/2}$ sums are listed in the rows following
  the corresponding main term. The total contribution from the core-excited
 autoionizing states are listed in rows labeled $\delta\alpha^{\text{core}}_S$
 and $\delta\beta^{\text{core}}_S$ of Tables~\ref{tab-comp1} and \ref{tab-comp2},
 respectively.

    We list the values of the reduced electric-dipole matrix elements and energies used
     in the calculation of $\alpha_S$
 and $\beta_S$ in Table~\ref{tab-comp1}. Since the same matrix elements and
 energies contribute to $\alpha_S$ and $\beta_S$, we do not repeat the
 energy and matrix element values in Table~\ref{tab-comp2}.
We use recommended values of energies from the National Institute of
Standards and Technology (NIST) database \cite{nist} when calculating all of the
polarizability values in this work.
The corresponding energy differences are listed in
columns two and three of Table~\ref{tab-comp1}.
Electric-dipole matrix elements evaluated using the SD
all-order method (Eq.~(\ref{eq18})) are given  in  columns labeled
$Z_{wv}$. It should be noted that the values of $Z_{wv}$
 given in Table~\ref{tab-comp1} generally differ by 0.4--1.0\%  from the
values of $Z_{wv}$ presented in
Ref.~\cite{tl-05}.  These differences arise  because we include the Breit interaction here
and, to the lesser extent, because of more accurate basis set used in the present work.

We also conducted a semiempirical scaling procedure, described for example,
in Refs.~\cite{Blundell,safr-alk}, for three transitions, $7S-7P_{1/2}$, $7S-7P_{3/2}$,
and $7P_{1/2}-6D_{3/2}$.
These transitions give significant contributions to the $\alpha_S$ and $\beta_S$
as well as Stark shifts that we discuss below. The scaling is carried out by
multiplying the values of the corresponding single valence excitation coefficients
by the ratio of the theoretical and experimental correlation energies and repeating the
calculation of the matrix elements with the modified excitation coefficients.
 Study of the breakdown of the correlation corrections showed that the dominant contribution to the
correlation correction to the values of these matrix elements comes from a single
term that contains only single valence excitation
coefficients. It has been shown that the scaling  works effectively in such
cases (see, for example \cite{Ca-new,Safronova}) since it is specifically aimed at correcting
the dominant contribution. In the case of the $7S-7P$ transitions, the correlation
breakdown is the same as in Cs, where the scaled values are in
excellent agreement with the high-precision experiment.
The scaling can not be applied to improve the values of the $6D_{3/2}-6P_{1/2}$
matrix element as the the corresponding correlation correction is not dominant,
leading to possibly reduced accuracy of this transition in comparison to
 $7S-7P_{1/2}$, $7S-7P_{3/2}$,
and $7P_{1/2}-6D_{3/2}$ ones.

\begin{table}
\caption{\label{tab-comp} The SD all-order Stark-induced scalar
and vector polarizabilities for the  $6P_{1/2} - 7P_{1/2}$
transition in Tl are compared with measurements
by \protect\citet{st-86} - $a$ and
 \protect\citet{st-94} - $b$.}
\begin{ruledtabular}
\begin{tabular}{lll}
\multicolumn{1}{c}{}& \multicolumn{1}{c}{$\alpha_S$}&
\multicolumn{1}{c}{$\left|\beta_S\right|$}\\
\hline
This work         &368       & 298     \\
Expt.$^{a}$ &247$\pm$12&198$\pm$10\\
Expt.$^{b}$ &377$\pm$8 &313$\pm$8\\
\end{tabular}
\end{ruledtabular}
\end{table}

We find that the contribution of the $6D_{3/2}$ term from Eqs.~(\ref{eq1}, \ref{eq12})
dominates the values of $\alpha_S$
and $\beta_S$. While all valence terms contributing to  $\alpha_S$
have the same sign, there is
significant cancellation of terms contributing to $\beta_S$. In fact, all $nS$
terms contribute with the sign opposite to that of the $nD_{3/2}$ terms with the
exception of the first ones, $7S$ and $6D_{3/2}$. Owing to this
cancellation, our value of $\beta_S$ may be somewhat less
accurate  than the value of $\alpha_S$.

In Table~\ref{tab-comp}, we compare our results for the
Stark-induced scalar and vector
polarizabilities for the $6P_{1/2} - 7P_{1/2}$ transition in Tl with experimental
measurements by \citet{st-94} and \citet{st-86}.
The conversion  factor between the units of $(\mu_B/c)$~(cm/V)  used in Refs.~\cite{st-94,st-86}
and atomic units used in the present work is $10^{-2}\alpha E_h/2e a_0$=1.8762$\times$10$^7$,
where $E_h$ is Hartree energy.
Our results support the measurements
of \citet{st-94} and clearly disagree with the measurements reported by
\protect\citet{st-86}. Our value of $\alpha_S$ is nearly within the
experimental uncertainty of the Ref.~\cite{st-94} measurement and our value of $\beta_S$ differs from the
central experimental value of Ref.~\cite{st-94} by 2$\sigma$. As we discussed above, accurate calculation of
$\beta_S$ is more difficult owing to the cancellation of the $nS$ and $nD_{3/2}$
contributions.

\section{Stark shift within the
$6P_{1/2}\ -7S_{1/2}$ and  $6P_{1/2}\ -7P_{1/2}$  transitions in
atomic thallium}

We calculate the Stark shifts within the $6P_{1/2} -7S_{1/2}$ and
$6P_{1/2} -7P_{1/2}$  transitions as a differences of scalar dipole
polarizabilities $\alpha$ of the $6P_{1/2}$ ground state and the
$7S_{1/2}$ or  $7P_{1/2}$ excited states. The expression for
$\alpha$ is given by (see, for example,
Refs.~\cite{st-70,mar-pol-04}):

\begin{eqnarray*}
\alpha(n_{0}P_{1/2})&=&\sum_{n}\left[ I_{S}(nS_{1/2})+I_{S}(nD_{3/2})\right]\\
\alpha(n_{0}S_{1/2})&=&\sum_{n}\left[ I_{S}(nP_{1/2})+I_{S}(nP_{3/2})\right],
\end{eqnarray*}
where
\[
I_{S}(nlj)=\frac{1}{3}\frac{Z_{n_{0}l_{0}j_{0},nlj}^{2}}
{E_{nlj}-E_{n_{0}l_{0}j_{0}}}.
\]
 Our results for the $6P_{1/2}$, $7P_{1/2}$, and $7S_{1/2}$ polarizabilities
 are given in Tables~\ref{tab-comp6} and ~\ref{tab-comp61}
where we use the same designations as in the previous section.
The polarizability $\alpha_{\rm core}$ of the Hg-like ionic core is
also evaluated using RPA approximation.
A more detailed discussion for the  $\alpha_{\rm core}$ in Na, K, Rb, Cs, and Fr
atomic systems is found in
Ref.~\cite{safr-alk}. We note that contributions from $\alpha_{\rm core}$
cancel when we evaluate the Stark shift  for a transition.
\begin{table}
\caption{\label{tab-comp6}  The contributions to the scalar dipole
 $6P_{1/2}$  and $7P_{1/2}$ polarizabilities $\alpha$(a.u.) in Tl.}
\begin{ruledtabular}
\begin{tabular}{lrrr}
\multicolumn{1}{c}{Contribution}&
\multicolumn{1}{c}{$nlj$}& \multicolumn{1}{c}{$\alpha(6P_{1/2})$ }&
\multicolumn{1}{c}{$\alpha(7P_{1/2})$\vspace{0.1cm}}\\
\hline
 $\alpha^{\text{main}}(nS)$       &  $7S_{1/2}$ &  9.2  &  -345  \\
                                  &  $8S_{1/2}$ &  0.5  &   616    \\
                                  &  $9S_{1/2}$ &  0.2  &    13    \\
                                  & $10S_{1/2}$ &  0.1  &     3   \\
 $\alpha^{\text{tail}}(nS)$       &             &  0.7  &     6  \\[0.5pc]
 $\alpha^{\text{main}}(nD_{3/2})$ & $6D_{3/2}$  & 11.0  &  4280  \\
                                  & $7D_{3/2}$  &  2.1  &   217  \\
                                  & $8D_{3/2}$  &  0.7  &    39  \\
                                  & $9D_{3/2}$  &  0.4  &    14  \\
  $\alpha^{\text{tail}}(nD_{3/2})$&             &  4.7  &    47  \\[0.5pc]
  $\alpha^{\text{core}}$          &             &  24.1  &    24   \\[0.2pc]
  $\delta\alpha^{\text{core}}_S$  &             &  -3.3  &    0    \\[0.2pc]
  Total                           &             &  50.4  &  4915   \\
\end{tabular}
\end{ruledtabular}
\end{table}

\begin{table}
\caption{\label{tab-comp61} The contributions to the scalar dipole
 $7S$ polarizability $\alpha$ in Tl. The energy differences
and absolute values of the electric-dipole reduced matrix elements for
relevant transitions are also listed. All values are given in a.u. }
\begin{ruledtabular}
\begin{tabular}{lrrrr}
\multicolumn{1}{c}{Contribution}&
\multicolumn{1}{c}{$nlj$}&
\multicolumn{1}{c}{$E_{nlj}-E_{7S}$ }& \multicolumn{1}{c}{$\left|Z_{7S,nlj}\right|$}&
\multicolumn{1}{c}{$\alpha(7S_{1/2})$\vspace{0.1cm}}\\
\hline
 $\alpha^{\text{main}}(nP_{1/2})$ &  $6P_{1/2}$ & -0.120640  &  1.826 &    -9.2    \\
                                  &  $7P_{1/2}$ &  0.035004  &  6.016 &   344.7      \\
                                  &  $8P_{1/2}$ &  0.067847  &  0.706 &     2.5      \\
                                  &  $9P_{1/2}$ &  0.081574  &  0.296 &     0.4     \\
 $\alpha^{\text{tail}}(nP_{1/2})$ &             &            &        &    0.4    \\[0.5pc]
 $\alpha^{\text{main}}(nP_{3/2})$ & $6P_{3/2}$  & -0.085134  & 3.397  &   -45.2   \\
                                  & $7P_{3/2}$  &  0.039565  & 8.063  &   547.7    \\
                                  & $8P_{3/2}$  &  0.069545  & 1.474  &    10.4    \\
                                  & $9P_{3/2}$  &  0.082401  & 0.713  &     2.1    \\
  $\alpha^{\text{tail}}(nP_{3/2})$&             &            &        &    3.0    \\[0.5pc]
  $\alpha^{\text{core}}$          &             &            &        &    24.1     \\[0.2pc]
  Total                           &             &            &        &    880.8       \\
\end{tabular}
\end{ruledtabular}
\end{table}

Only the polarizability contributions are listed in Table~\ref{tab-comp6},
since all relevant energies and electric-dipole matrix elements are
already listed in Table~\ref{tab-comp1}. The calculation of the
$7S_{1/2}$ polarizabilities involves the calculation of other
series of the matrix elements, $7S-nP_{1/2}$ and $7S-nP_{3/2}$.
We list those values, calculated using SD all-order method,
together with the corresponding energy differences taken from
\cite{nist} in Table~\ref{tab-comp61}.

The value of the $6P_{1/2}$ polarizability has two dominant (and nearly
equal) contributions, from $6P_{1/2}-6D_{3/2}$ and $6P_{1/2}-7S$
transitions. The value of the $7P_{1/2}$ polarizability is  dominated
by the contribution from $7P_{1/2}-6D_{3/2}$
transition. The value of the $7S$ polarizability is  dominated
entirely by the contributions from both $7S-7P$
transitions. As we have discussed above, we conducted more
accurate calculation of the $7P_{1/2}-6D_{3/2}$, $7S-7P_{1/2}$,
and  $7S-7P_{3/2}$ electric-dipole matrix elements by rescaling
the single valence excitation coefficients with the correct value of the
correlation energy leading to more accurate evaluation of the dominant
contributions to $7S$ and $7P_{1/2}$ polarizabilities.
Our value for the  $\alpha(6P_{1/2})$ =
50.4~a$_0$$^3$ is in good agreement with the theoretical result
49.2~a$_0$$^3$ by \citet{kozlov}.

\begin{table}
\caption{\label{tab-comp7} The all-order values of the ground and excited
state polarizability differences are compared with experimental
 results by \protect\citet{st-02} - $a$,
    by \protect\citet{st-94}- $b$, and
 by \protect\citet{st-70}- $c$.  }
\begin{ruledtabular}
\begin{tabular}{lll}
\multicolumn{1}{c}{}& \multicolumn{1}{c}{$\alpha(6P_{1/2})\ - \alpha(7S_{1/2})$}&
\multicolumn{1}{c}{$\alpha(6P_{1/2})\ - \alpha(7P_{1/2})$}\\
\hline
Present work         &-830             & -4866     \\
Expt.$^{a}$ &-829.7$\pm$3.1       &            \\
Expt.$^{b}$ &-900$\pm$48      &-4967$\pm$249\\
Expt.$^{c}$ &-776$\pm$80       &            \\
\end{tabular}
\end{ruledtabular}
\end{table}

In Table~\ref{tab-comp7}, we compare our results for  the Stark shift in the
$6P_{1/2} -7S_{1/2}$ and $6P_{1/2}\ -7P_{1/2}$ transitions
 with experimental results from Refs.~\cite{st-02,st-94,st-70}.
The conversion factor between the $\Delta \nu_S$ in kHz/(kV/cm)$^2$ units used in Refs.~\cite{st-02,st-94,st-70}
to polarizabilities in atomic units used in the present work is
$2\times 10^{-7} h/(4 \pi \epsilon_0 a_0^3)=8.03756$, where $h$ is the Plank constant.
Our result for the $6P_{1/2} - 7S_{1/2}$ Stark shift agrees with the most accurate, 0.4\%,
experimental value from Ref.~\cite{st-02}
within the experimental uncertainty. The all-order value of the Stark shift for the $6P_{1/2}\ -7P_{1/2}$ transition
also agrees with the experiment within the experimental uncertainty.
We note that this comparison essentially
tests the accuracy of the $7S$ and $7P_{1/2}$  polarizability
calculations since the ground state polarizability is small
comparing to the excited state polarizabilities.
\section{Conclusion}
In summary, a systematic  study using relativistic SD all-order method of
the Stark
shift within the $6P_{1/2} -7S_{1/2}$ and  $6P_{1/2} -7P_{1/2}$
transitions and the  Stark-induced amplitudes
$\alpha_S$ and $\beta_S$ in the $6P_{1/2} - 7P_{1/2}$ transition in
atomic thallium is presented. Our results $\alpha$ =
368~a$_0$$^3$ and  $\left|\beta_S\right|$= 298~a$_0$$^3$ support the experimental
measurements carried out by D. DeMille, D. Budker, and E. D. Commins [Phys.\
Rev.\ A{\bf 50}, 4657 (1994)].  Our results for Stark shifts
in the $6P_{1/2} -7S_{1/2}$ and  $6P_{1/2} -7P_{1/2}$ transitions
 agree with the most accurate measured values
within the experimental uncertainties for both transitions.

\begin{acknowledgments}
   The work was supported in part by DOE-NNSA/NV
Cooperative Agreement DE-FC52-01NV14050. The work of MSS was
supported in part by National Science Foundation Grant  No.\
PHY-04-57078. The work of WRJ was supported in part by NSF Grant
No.\ PHY-04-56828.
\end{acknowledgments}

\end{document}